# MANGO: Multimodal Acuity traNsformer for intelliGent ICU Outcomes

**Jiaqing Zhang[1,2], Miguel Contreras[2,3], Sabyasachi Bandyopadhyay[4], Andrea Davidson[2,5], Jessica Sena[2,3], Yuanfang Ren[2,5], Ziyuan Guan[2,5], Tezcan Ozrazgat-Baslanti[2,5], Tyler J. Loftus[6], Subhash Nerella[2,3], Azra Bihorac[2,5], Parisa Rashidi[2,3]***

[1]Department of Electrical and Computer Engineering, University of Florida, Gainesville, FL, USA

[2]Intelligent Clinical Care Center (IC3), University of Florida, Gainesville, FL, USA

[3]Department of Biomedical Engineering, University of Florida, Gainesville, FL, USA

[4]Department of Medicine, Stanford University, Stanford, USA

[5]Division of Nephrology, Department of Medicine, University of Florida, Gainesville, FL, USA

[6]Department of Surgery, University of Florida, Gainesville, FL, USA

**\* Correspondence:**

**Parisa Rashidi**

parisa.rashidi@ufl.edu

# Abstract

Estimation of patient acuity in the Intensive Care Unit (ICU) is vital to ensure timely and appropriate interventions. Advances in artificial intelligence (AI) technologies have significantly improved the accuracy of acuity predictions. However, prior studies using machine learning for acuity prediction have predominantly relied on electronic health records (EHR) data, often overlooking other critical aspects of ICU stay, such as patient mobility, environmental factors, and facial cues indicating pain or agitation. To address this gap, we present MANGO: the Multimodal

Acuity traNsformer for intelliGent ICU Outcomes, designed to enhance the prediction of patient acuity states, transitions, and the need for life-sustaining therapy. We collected a multimodal dataset ICU-Multimodal, incorporating four key modalities: EHR data, wearable sensor data, video of patient's facial cues, and ambient sensor data, which we utilized to train MANGO. The MANGO model employs a multimodal feature fusion network powered by Transformer masked self-attention method, enabling it to capture and learn complex interactions across these diverse data modalities even when some modalities are absent. Our results demonstrated that integrating multiple modalities significantly improved the model's ability to predict acuity status, transitions, and the need for life-sustaining therapy. The best-performing models achieved an area under the receiver operating characteristic curve (AUROC) of 0.76 (95% CI: 0.72–0.79) for predicting transitions in acuity status and the need for life-sustaining therapy, while 0.82 (95% CI: 0.69–0.89) for acuity status prediction. This study is the first to incorporate all four modalities in predicting ICU patient outcomes. Our findings highlight MANGO's potential to enhance patient monitoring in the ICU through advanced multimodal data integration, offering a powerful tool for improving clinical decision-making.

## I. Introduction

The Intensive Care Unit (ICU) is a highly complex medical environment where multiple factors, such as patient condition severity, treatment effectiveness, clinician expertise, and environmental conditions, collectively influence patient outcomes. Estimating the acuity of the patient under such a complex condition is tricky, given acuity levels can fluctuate significantly and frequently during an ICU stay. Timely and accurate estimation with taking more comprehensive consideration of multiple aspects in the ICU is crucial for improving survival rates and supporting recovery [1].

Rapid advancements in artificial intelligence (AI) have transformed healthcare, enabling more efficient and accurate patient assessments [2]. Multiple studies have used neural networks for patient acuity assessment [3, 4]. However, these studies are limited to EHR data and do not include the other aspects of an ICU stay, such as patient mobility, facial cues related to pain or agitation, and ICU environmental information including light intensity and sound pressure level related to sleep quality [5]. Multiple prior works have demonstrated associations between these data modalities and patient acuity [6, 7], highlighting their capacity to benefit clinical outcome prediction.

Multimodal models have emerged as a powerful tool for integrating diverse data sources in complicated setups [8, 9]. By combining multiple data types, multimodal models improve diagnostic accuracy, prognostic predictions, and personalized treatment planning. For example, Ma et al. [10] introduced a contrastive learning-based multimodal model that utilized EHR data and clinical notes, achieving state-of-the-art performance in predicting nine postoperative complications. Similarly, Li et al. [11] developed the eXplainable Multimodal Mortality Predictor (X-MMP), which integrated clinical notes, discrete event sequences, and vital signs. This approach outperformed traditional single-modality models in predicting mortality, highlighting the potential of multimodal systems to surpass existing standards in healthcare analytics.

In this paper, we propose a **M**ultimodal **A**cuity tra**N**sformer for intelli**G**ent ICU **O**utcomes model (**MANGO**) trained on a multimodal dataset, ICU-Multimodal. ICU-Multimodal was collected at the University of Florida (UF) Health Shands Hospital, consisting of four modalities from 310 patients: structured EHR data, wearable sensor data, patient facial action units (AUs) video data, and ambient sensor data. Our experiments with different combinations of the modalities demonstrated the robustness and efficacy of the prediction on two tasks: 1) transitions in the acuity status and the need for life-sustaining therapies and 2) patient acuity status. Notably, the model trained with EHR data and the other three modalities exhibited the best overall performance on transition prediction, achieving an Area Under the Receiver Operating Characteristic Curve (AUROC) of 0.76. In comparison, the model trained on EHR data only, which we defined as the baseline model, achieved an AUROC of 0.71. In the case of the acuity status classification task, we achieved an AUROC of 0.82 with MANGO trained on all four mobilities, compared with an AUROC of 0.70 when using the baseline model. Further, we computed the integrated gradient attributions for each feature to capture the importance across the four modalities. To the best of our knowledge, MANGO is the first multimodal model involving four different modalities in predicting ICU outcomes.

The key contributions of our work are summarized as follows:

1) We propose a novel, robust, and efficient multimodal model, MANGO, which integrates four different data modalities to predict acuity status, transitions in acuity status, and transitions of the need for life-sustaining therapies in the ICU, as shown in Fig. 1.

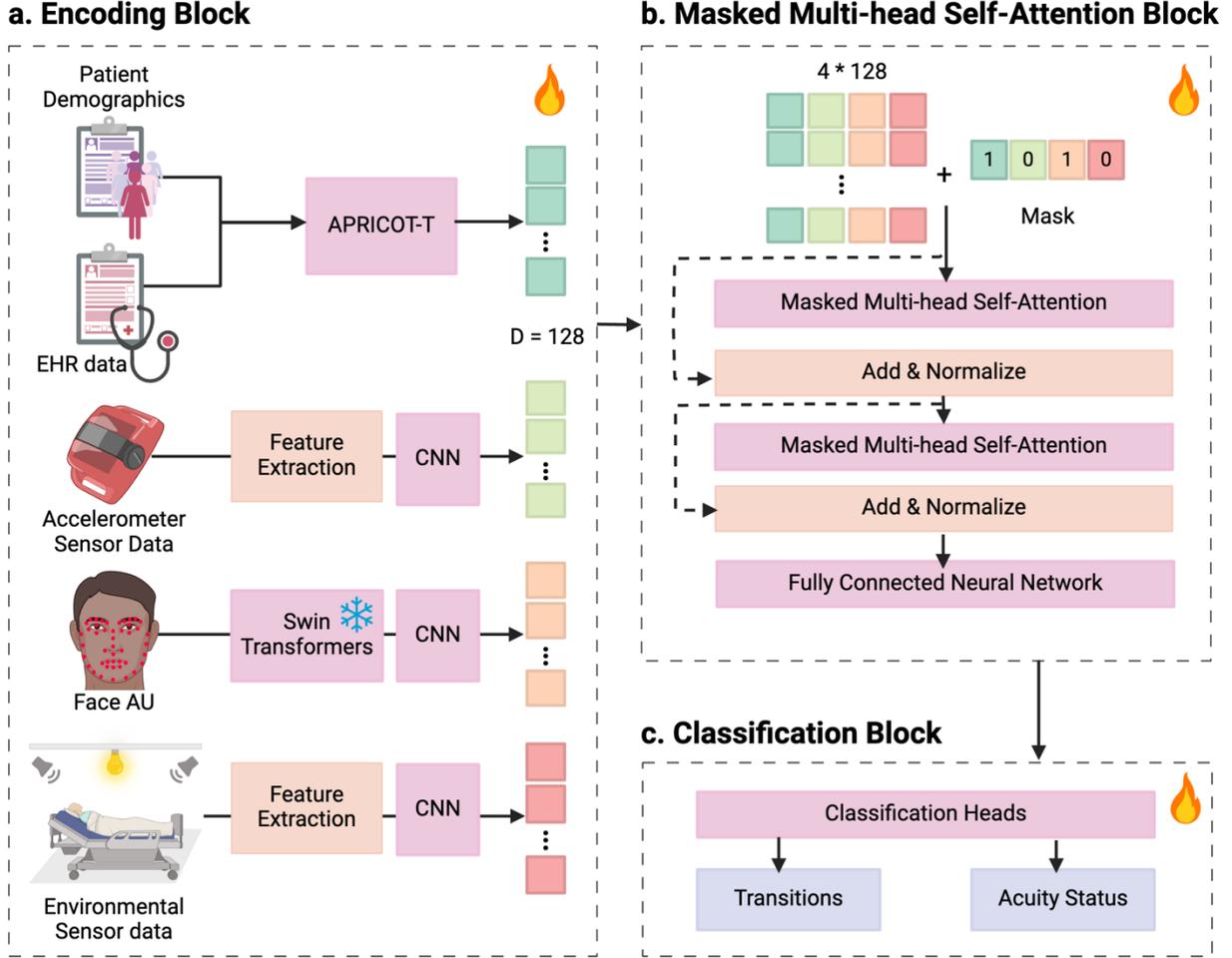

Fig. 1. The workflow of the proposed model. We employed different models to encode the respective modalities into four $D = 128$ vectors. While training, the Swin Transformer weights were frozen, and other blocks were trainable. These vectors were then stacked into an $4 \times 128$ sequence, where the sequence length is 4, and the dimension of the embeddings is 128. Along with the mask, this sequence was fed into two multi-head self-attention blocks, each equipped with residual connections and layer normalization. The resulting attention score was processed through a fully connected neural network. Finally, the classification heads addressed two distinct tasks, i.e., acuity status and transitions.

2) We designed a novel feature fusion strategy that makes each modality optional so that our pipeline can be easily translated to clinical care and be helpful in analyzing incomplete datasets directly.

# II. Methodology

**Notations Definition**. The multimodal dataset ICU-Multimodal $D$ consists of $N = 310$ patients. For each patient $p_i$, we ensured the presence of EHR data ($X_{EHR}$), beyond which we collected wearable sensor data in the form of accelerometer data ($X_{Accel}$), patient facial action units (AUs) extracted from red-green-blue (RGB) video data ($X_{Face}$), and ambient sensor data of environmental factors ($X_{Env}$) including light intensity and noise levels in the ICU room. For the data in each modality set $X$, we split them into 4-hour intervals, which we will refer to as the observation window. We refer to the 4-hour window following the observation window as the prediction window. Based on the presence and absence of the modality set $X$, we created an attention mask ($M$) to indicate the presence and absence of individual data modalities for each patient within each observation window. The resulting ICU-Multimodal dataset $D = \{p^i\,(X_{EHR}^i, X_{Accel}^i, X_{Face}^i, X_{Env}^i), M^i\}^N$ was then processed and provided to the MANGO model to predict transitions ($Y_{trans}$) of the acuity status and life-sustaining therapies and acuity status ($Y_{status}$) within the prediction window.

## A. Data Collection and Processing

**1) Data sources:** The ICU-Multimodal dataset was collected from adult patients who provided informed consent to this research study during their admission to one of nine specialized ICUs at the UF Health Shands Hospital main campus in Gainesville, Florida. The study was approved by the University of Florida Institutional Review Board under IRB201900354 and IRB202101013 and was conducted in compliance with all relevant federal, state, and university laws and regulations.

**2) Multimodal dataset collection:** Participants consented to collect their prior medical history, entire hospital admission, complications, and mortality data for up to five years after their involvement in the study. The UF Integrated Data Repository provided these data with de-identified patient IDs.

Data collection for each patient took place for 7 days, or until they were transferred or discharged from the ICU. Accelerometry data was collected using Shimmer ECG (Shimmer Sensing, Dublin, Ireland) and Actigraph GTX3+ devices (ActiGraph LLC, Pensacola, FL, USA), which patients wore on their wrists and/or ankles for the duration of the study. We requested that the nursing staff document instances of device removal (i.e., removal for surgery or bathing), and known removal and reapplication times were noted as device downtimes and excluded from the study analysis. The RGB video data was recorded using an Amcrest camera model IP2M-841W-V3 (Amcrest

TABLE I
MODALITY AND CLASS DISTRIBUTION AT OBSERVATION WINDOWS LEVEL

| Item | Development* | Test | p-value** |
|---|---|---|---|
| **Modalities, n (%)** | | | |
| EHR | 33,779 (100.0%) | 8,349 (100.0%) | 1.00 |
| Facial AU features | 3,741 (11.0%) | 877 (10.5%) | 0.14 |
| Accelerometer features | 3,634 (6.8%) | 634 (7.6%) | < 0.05 |
| Environmental features | 4,683 (13.9%) | 1228 (14.7%) | 0.05 |
| **Classes, n (%)** | | | |
| Stable to unstable | 423 (1.2%) | 85 (1.1%) | 0.08 |
| Unstable to stable | 331 (1.0%) | 59 (0.7%) | < 0.05 |
| MV to no MV | 419 (1.2%) | 88 (1.0%) | 0.16 |
| No MV to MV | 325 (1.0%) | 64 (0.8%) | 0.09 |
| VP to no VP | 209 (0.6%) | 29 (0.3%) | < 0.05 |
| No VP to VP | 191 (0.6%) | 26 (0.3%) | < 0.05 |
| Discharge | 84 (0.2%) | 22 (0.3%) | 0.81 |
| Stable | 22,269 (65.9%) | 6,691 (80.1%) | <0.05 |
| Unstable | 11,412 (33.8%) | 1,633 (19.6%) | <0.05 |
| Deceased | 14 (0.0%) | 3 (0.0%) | 0.82 |

Abbreviations: n: number, AUs: action units, MV: Mechanical ventilation, VP: vasopressors.
*The development set is further divided into training and validation subsets in a 9:1 ratio.
**P-values are calculated by a two-proportion z-test.

Industries LLC, Houston, TX, USA), zoomed directly at the patient's face, and recorded at 30 frames per second. The camera is mounted on a standalone cart with a computer and monitor. A custom user interface was developed to control the data collection process and allow ICU coordinators to pause the recording at any time. The facial AUs were manually annotated on the frames extracted from RGB video data by annotators trained in a facial action coding system (FACS) [12]. The ambient sensor data was recorded using Actigraph GTX3+ device, Apple iPod (Apple Inc., Cupertino, CA, USA), and Thunderboard Sense 2 (Silicon Labs, Austin, TX, USA).

Throughout the study period, the clinical research team conducted daily visits to ensure the study equipment was correctly positioned and functioning properly. They documented any instances of device downtime and restarted recordings as necessary to maintain consistent data collection.

*3) Data processing:* All sensor data were stored locally on an encrypted and password-protected hard drive and then manually uploaded to a server through a secured VPN connection. The raw data collected was curated by clinical staff to remove protected health information and stored in patient-specific folders. Multiple data pipelines were developed for each modality to automate data preparation for annotation and model training. Docker containers were utilized to manage services and pipelines running on the server in a standalone manner. The data curation and post-processing were performed on a secured server with access restrictions and connectivity available only through a health VPN.

*4) Development & Test sets splitting:* The processed data were split into development and test sets based on patients' unique ID by the ratio 8:2. The development set was then further split into training and validation sets by 9:1. No patient overlap between train, validation, and test sets was present to avoid data leakage. For each modality set $X$, the data was split into 4-hour observation windows, starting at $p_i$'s admission. We obtained 33,779 observation windows in the development set and 8,349 in the test set. The distribution of the modalities at the observation window level is shown in Table I.

*5) Prediction Outcomes:* We generated acuity labels, stable and unstable, for every four-hour prediction window based on the criteria proposed by Ren et al. [13] for developing computable phenotypes for acuity status. Patients are considered unstable if they require at least one of the four life-supportive therapies (mechanical ventilation, MV; massive blood transfusion, BT; vasopressor, VP; continuous renal replacement therapy, CRRT); if not, they are considered stable. Life-sustaining therapies in this study include administering MV and VP drugs to the patient. Based on the acuity status labels and life-sustaining therapies labels, we further generated the following transition labels: transitions in acuity status included patient status changing from stable to unstable and unstable to stable. Similarly, transitions in therapies included "MV to no MV", "no MV to MV", "VP to no VP", and "no VP to VP". For $Y_{status}$, we generated labels including "discharge", "stable", "unstable", and "deceased". The "discharge" and the "deceased" label were extracted from the patient's admission information. The distributions of the labels are shown in Table I.

## B. Multimodality feature fusion

We designed a mid-level multimodal feature fusion approach to combine different modalities of varying together. We encoded the raw features of each modality into a single vector with a consistent embedding dimension of 128. Then, we concatenated the four vectors as a sequence with four elements: EHR data $X_{EHR}$, accelerometry data $X_{Accel}$, face data $X_{Face}$, and environmental data $X_{Env}$. Each vector stored the information of the modality recorded in four-hour observation window. We also generated a corresponding sequence mask for each observation

## TABLE II
### FEATURES FOR ACCELEROMETER DATA

| Features | Description |
|---|---|
| MVM | Mean of vector magnitude |
| SDVM | SD of vector magnitude |
| MANGLE | Mean of angle between x and vector magnitude |
| SDANGLE | SD of angle between x and vector magnitude |
| DF | Dominant frequency |

Abbreviations: SD: standard deviation.

## TABLE III
### FEATURES FOR FACE AUs DATA

| Features | Description |
|---|---|
| AU1 | Inner Brow Raiser |
| AU2 | Outer Brow Raiser |
| AU6 | Cheek Raiser |
| AU7 | Lid Tightener |
| AU10 | Upper Lip Raiser |
| AU12 | Lip Corner Puller |
| AU25 | Lips Part |
| AU26 | Jaw Drop |
| AU43 | Eyes Close |

Abbreviations: AUs: action units.

window to account for the missing data modalities in the sample, as shown in Fig. 1.

**EHR data.** Previously, the APRICOT-T model [14] was developed and validated to predict acuity state, transitions between acuity states, and the need for life-sustaining therapies using EHR data only, outperforming clinical baseline and other machine learning and deep learning baselines. We followed the same data preprocessing strategy in this work on EHR data. We included vital signs, laboratory results, assessment scores, and medications to generate temporal variables within four-hour observation windows. We extracted static patient information (age, sex, race, and comorbidities) from admission and combined it with temporal variables data as the EHR component $X_{EHR}$.

**Accelerometer data.** The accelerometer data collected using multiple devices was downsampled to a consistent 10 Hz sampling frequency to ensure uniformity in the input data rate. Following Kheirkhahan et al.'s work [15], we extracted statistical features based on raw data over the four-hour observation window, as shown in Table II. These features have been used widely to quantify physical activity type, intensity, and energy expenditure. Vector magnitude reflects the force of the patient's movement, while the angle between x and vector magnitude represents the orientation of the movement. Other than time domain information, we included the dominant frequency to show the frequency of the patient's movement. Additionally, we included "position" as an additional binary feature since the accelerometer data was collected on either the ankle or wrist of the patients. If patients had both data, we kept wrist.

**Face AUs data.** Facial expressions are related to pain levels and agitation, which can provide additional information for acuity-level assessments [7]. We employed the SWIN transformer model [16] trained on multiple datasets: BP4D [17], DISFA [18], UNBC [19], and UF AU-ICU [20] to generate the Face AUs prediction for multimodal model training [7]. We use a fine-tuned YOLO model [21] to detect and crop patients' faces in the RGB videos collected from the ICU. Subsequently, the SWIN transformer model was applied to predict 9 facial AUs for each sample, as detailed in Table III. We then calculated the percentage presence of each AU over four-hour observation periods as the face component ($X_{Face}$) of the sequence.

**Environmental data.** Inspired by the work of Bandyopadhyay et al. [22], which utilized a deep learning model to predict the risk of delirium based on environmental factors (light intensity and sound pressure levels), we incorporated ambient variables into our multimodal model. Specifically, we included the average light intensity (l), the minimum and maximum sound pressure levels (Lmin & Lmax), and the average sound pressure level of four-hour observation windows (Lmean), which had been proven correlated to delirium prediction as the environmental element. The environmental factors can affect a patient's circadian rhythms and sleep quality, thereby influencing their recovery [23].

**Feature fusion.** After obtaining the four feature modality elements, we normalized the values of each element to (0, 1). Then, we utilized four different encoders to generate the embeddings for each modality element. For the EHR element, APRICOT-T [3] was used to generate the embeddings. Another three isolated convolutional neural networks (CNN) were used to generate embeddings of the other three modalities, as shown in the Encoding Block of Fig. 1. The CNN blocks consisted of two 1D convolutional layers with ReLU activations, followed by a flattening layer and a fully connected linear layer. During the training, the weights of APRICOT-T encoder were initialized, and trainable together with the other three CNNs. The embeddings of each modality had a consistent dimension of 128.

Then, we combined the four modality embeddings as a sequence $p^i(X_{EHR}^i, X_{Face}^i, X_{Accel}^i, X_{Env}^i) \in \mathbb{R}^{N \times 4 \times 128}$ for each patient. An aligned mask $M \in \mathbb{R}^{N \times 4}$ was created based on the sequence to mask those modalities missing in each sequence. For example, if in $p^i$, $X^i_{Face} = NaN$, $X^i_{Accel} = NaN$, and $X^i_{EHR} \neq NaN$, $X^i_{Env} \neq NaN$, the mask is indicated as $M_i = (1,0,0,1)$.

## C. Masked Multi-Head Self-Attention Modeling

Masked multi-head self-attention (MMSA) is a key component in the Transformer architecture [24], particularly in tasks such as language modeling where the goal is to predict the next word in a sequence. Our bidirectional encoder architecture includes two MMSA blocks with residual connection and layer normalization to omit the influence of the missing modalities in the prediction and learn the context solely from existing modalities. The MMSA can be represented as:

$$MMSA(Q, K, V, M) = softmax(\frac{QK^T}{\sqrt{d_k}} + M)V \qquad (1)$$

Where $Q, K, V$ represent the query, key, and value matrices, respectively, with the embedding size of 128 and $d_k$ is the dimensionality of the key vectors. $M$ is the mask we generated based on the presence and absence of the modalities, where each element $M_i$ is set to $-\infty$ if the modality at position $i$ is missing or equals 0 in our case. The $-\infty$ entries in $M$ resulted in zero probabilities after the SoftMax activation in the attention computation.

After obtaining the attention score from our encoder model, a shared backbone 3-layer fully connected network was attached before the classification block to aid in integrating information across different attention heads, as shown in Fig. 1.

## D. Classification and Analysis

**1) Classification tasks definition:** We defined two types of classification tasks: 1) Transition classification: transitions in patient acuity status and need for life-sustaining therapies, as defined by the transition between stable and unstable states, MV and no MV, and VP and no VP. 2) Status classification: acuity state of the patient, i.e., discharge, stable, unstable, and deceased. Both tasks are needed to provide a comprehensive understanding of the patient's clinical trajectory, enabling precise and timely interventions. We have implemented six classification heads to predict the transitions from the output of our multimodal encoder and four heads for patient acuity status prediction. In total, ten classification heads were implemented after the MMSA block and the shared backbone network for the ten sub-tasks.

TABLE IV
PATIENT CHARACTERISTICS

| Characteristics | Development | Test | p-value |
|---|---|---|---|
| **Basic information** | | | |
| Number of patients, n | 248 | 62 | - |
| Age, mean (SD) | 66 (17) | 62 (17) | 0.16 |
| Female, n (%) | 90 (36%) | 21 (34%) | 0.72 |
| Length of stay (hours), median (25th, 75th percentile) | 218 (113, 487) | 179 (92, 401) | 0.42 |
| **Modalities, n (%)** | | | |
| Face AUs data | 213 (86%) | 56 (90%) | 0.36 |
| Accelerometer data | 147 (59%) | 33 (53%) | 0.39 |
| Environmental data | 187 (75%) | 48 (77%) | 0.74 |
| **Race, n (%)** | | | |
| Black | 31 (13%) | 11 (18%) | 0.28 |
| White | 200 (81%) | 48 (77%) | 0.57 |
| Other | 17 (7%) | 3 (5%) | 0.56 |
| **Comorbidities, n (%)** | | | |
| Cancer | 20 (8%) | 3 (5%) | 0.39 |
| Cerebrovascular disease | 28 (11%) | 7 (11%) | 0.99 |
| Dementia | 4 (2%) | 3 (5%) | 0.13 |
| Paraplegia hemiplegia | 16 (6%) | 8 (13%) | 0.09 |
| Congestive heart failure | 65 (26%) | 21 (34%) | 0.23 |
| Chronic obstructive pulmonary disease | 69 (28%) | 12 (19%) | 0.17 |
| Diabetes | 44 (18%) | 12 (19%) | 0.33 |
| Liver disease | 52 (21%) | 15 (24%) | 0.58 |
| Peptic ulcer | 8 (3%) | 2 (3%) | 0.99 |
| Renal disease | 67 (27%) | 17 (27%) | 0.95 |

Abbreviations: SD: standard deviation, n: number, AUs: action units.
P-values are calculated by Welch's t-test and two-proportion z-test for continuous variables and categorical variables appropriately

2) **Evaluation metrics:** We used AUROC to evaluate the classification performance. To determine if the classification performance difference between the baseline model (only EHR features) and other configurations that included at least one additional modality during training was statistically significant, we compared all metric values using the Wilcoxon rank-sum test. We performed a 100-iteration bootstrap to calculate the 95% confidence interval (CI) for each performance metric, with the median value across the bootstrap iterations representing the overall value for each metric.

3) **Interpretation:** We used the integrated gradients [29] technique to assess feature importance across various modalities to interpret the MANGO model predictions. For EHR embeddings, we computed the integrated gradient attributions for each temporal variable within the APRICOT-T encoder layers. For the remaining three modalities, we analyzed and compared the integrated gradient attributions across each CNN block.

## III. Results

### A. Patient Characteristics

A total of $N = 310$ patients were part of the ICU-Multimodal dataset, as shown in Table IV. The average age of the development set and the test set were 66 and 62, respectively. In both sets, a higher number of male patients was observed, with percentages of 66% and 64%. The white patients constituted the majority in both sets. The median length of stay in the development set was 39 hours longer than the test set, while no significant difference was observed. In the development set, 86% of patients had face AUs data, while 90% had this data in the test set. Fewer patients had accelerometer data compared to the other modalities, with 59% in the development set and 53% in the test set. For environmental data, 75% and 77% of patients had light and sound features recorded in the development and test sets, respectively. In terms of comorbidities, no significant difference was observed among the development and test sets based on the p-values.

### B. Experiments

**1) Experimental setups:** We conducted a comparative analysis of six modality combinations to assess model performance. The baseline model was trained exclusively on EHR data (EHR baseline). The advanced models utilized either all four modalities (All), or pairwise modalities, i.e., EHR in conjunction with each of the other singular modalities: EHR with accelerometer data (EHR + Accel), EHR with facial data (EHR + face), and EHR with environmental data (EHR + Env). Additionally, we examined the performance of the model trained on EHR combined with both accelerometer and facial data without environmental features (EHR + Accel + face).

An early-stopping strategy was implemented, with a patience threshold of 10 epochs. For each experimental setup, the optimal model was selected based on the highest AUROC across three critical tasks, i.e., stable to unstable, no MV to MV, and no VP to VP. All models were trained on an NVIDIA A100-SXM4-80GB GPU.

TABLE V
CLASSIFICATION PERFORMANCE – ACUITY STATE TRANSITIONS

| Outcomes | Features AUROC (95% CI) ↑ | | | | | |
|---|---|---|---|---|---|---|
| | EHR baseline | EHR+ Accel | EHR+ Face | EHR+ Env | EHR + Accel+ Face | All |
| Unstable-Stable | 0.74 (0.69-0.80) | 0.73 (0.67-0.78)* | 0.75 (0.69-0.79)* | 0.75 (0.70-0.81)* | 0.74 (0.69-0.80) | 0.74 (0.70-0.78) |
| Stable-Unstable | 0.59 (0.50-0.66) | 0.67 (0.59-0.76)** | 0.73 (0.65-0.80)** | 0.68 (0.62-0.75)** | 0.70 (0.63-0.78)** | 0.71 (0.64-0.79)** |
| MV-No MV | 0.72 (0.66-0.76) | 0.70 (0.66-0.76)** | 0.75 (0.70-0.79)** | 0.75 (0.70-0.81)** | 0.74 (0.69-0.79)** | 0.74 (0.71-0.78)** |
| No MV-MV | 0.62 (0.53-0.71) | 0.66 (0.59-0.75)** | 0.73 (0.67-0.81)** | 0.69 (0.63-0.76)** | 0.71 (0.65-0.79)** | 0.72 (0.67-0.79)** |
| VP-No VP | 0.84 (0.78-0.90) | 0.83 (0.77-0.88)* | 0.80 (0.74-0.87)** | 0.81 (0.71-0.87)** | 0.87 (0.82-0.91)** | 0.86 (0.80-0.90)** |
| No VP-VP | 0.76 (0.62-0.84) | 0.68 (0.55-0.81)** | 0.75 (0.66-0.84)* | 0.75 (0.66-0.84)* | 0.75 (0.62-0.85)* | 0.76 (0.65-0.86) |
| **Overall** | 0.71 (0.67-0.75) | 0.71 (0.68-0.76) | 0.75 (0.72-0.78)** | 0.74 (0.70-0.77)** | 0.75 (0.72-0.79)** | **0.76 (0.72-0.79)** |

Abbreviations: CI: Confidence interval, Accel: Accelerometer, Env: Environmental, MV: Mechanical ventilation, VP: vasopressors.
Color coding: yellow: p-value > 0.05 compared to baseline, red: p-value < 0.05 compared to baseline and the performance dropped, green: p-value < 0.05 compared to baseline and the performance boosted.
The best performance of each outcome is bolded.
P-values are based on pairwise Wilcoxon rank sum tests.
*: p-value < 0.05 and p-value > 0.001 compared to baseline.
**: p-value < 0.001 compared to baseline.

2) **Experiment results:** We evaluated the performance across six different experimental setups of the different combinations of modalities (EHR baseline, EHR + Accel, EHR+ Face, EHR + Env, EHR + Accel + Env, and All), as shown in Table V and Table VI.

For predicting transitions in acuity status and life-sustaining therapies, the model trained on all four modalities achieved the best overall performance, with an AUROC of 0.76 (0.72-0.79 95% CI). Notably, it outperformed the baseline in all the classes except for "No VP to VP". For the models trained on pairwise modalities, the model trained on EHR + Accel data did not outperform the baseline, with an AUROC of 0.71 (0.68-0.76 95% CI), while the models trained on EHR + Face data and the model trained on EHR + Env data showed statistically superior classification performance than the baseline. The EHR + Accel + face model showed comparable performance as the model trained on all four modalities, with an AUROC of 0.75 (0.72-0.79 95% CI).

TABLE VI
CLASSIFICATION PERFORMANCE: ACUITY STATUS

| Outcomes | Features AUROC (95% CI) ↑ | | | | | |
|---|---|---|---|---|---|---|
| | EHR baseline | EHR+ Accel | EHR+ Face | EHR+ Env | EHR+ Accel+ Face | All |
| Discharge | 0.81 (0.72-0.89) | 0.81 (0.72-0.88) | 0.80 (0.67-0.87) | 0.77 (0.70-0.86)** | 0.82 (0.75-0.88) | 0.79 (0.70-0.88)* |
| Stable | 0.73 (0.72-0.74) | 0.79 (0.78-0.80)** | 0.86 (0.85-0.87)** | 0.85 (0.84-0.86)** | 0.85 (0.84-0.86)** | 0.86 (0.85-0.86)** |
| Unstable | 0.82 (0.81-0.83) | 0.80 (0.79-0.81) | 0.87 (0.86-0.88)** | 0.86 (0.85-0.86)** | 0.86 (0.85-0.87)** | 0.86 (0.85-0.87)** |
| Deceased | 0.43 (0.12-1.00) | 0.88 (0.70-1.00)** | 0.49 (0.05-1.00) | 0.42 (0.01-1.00) | 0.40 (0.02-1.00) | 0.82 (0.26-1.00)** |
| **Overall** | 0.70 (0.61-0.85) | 0.82 (0.78-0.86)** | 0.76 (0.64-0.89)** | 0.73 (0.62-0.88)* | 0.74 (0.64-0.89)** | **0.82 (0.69-0.89)** |

Abbreviations: CI: Confidence interval, Accel: Accelerometer, Env: Environmental, MV: Mechanical ventilation, VP: vasopressors.
Color coding: yellow: p-value > 0.05 compared to baseline, red: p-value < 0.05 compared to baseline and the performance dropped, green: p-value < 0.05 compared to baseline and the performance boosted.
The best performance of each outcome is bolded.
P-values are based on pairwise Wilcoxon rank sum tests.
*: p-value < 0.05 and p-value > 0.001 compared to baseline.
**: p-value < 0.001 compared to baseline.

In the acuity status prediction task (Table VI), the model trained on all four modalities and EHR + Accel achieved the best performance with AUROCs of 0.82 (0.69-0.89) and 0.82 (0.78-0.86), respectively. They showed robust classification performance in mortality prediction with a highly imbalanced label distribution, making it a challenging task. The model trained on all modalities did not boost the "Discharge" prediction performance.

## C. Feature importance

The integrated gradient analysis [25] shown in Fig. 2 illustrates the contribution level of different features from the four different modalities. In both tasks, EHR features contributed the most to the prediction. The other three modalities had similar trends of feature importance. The mean of the angle between x and vector magnitude and AU43 was the most important feature in transition prediction and status classification. The sound pressure level ranked highly for the environmental data, but the light intensity only showed limited contributions.

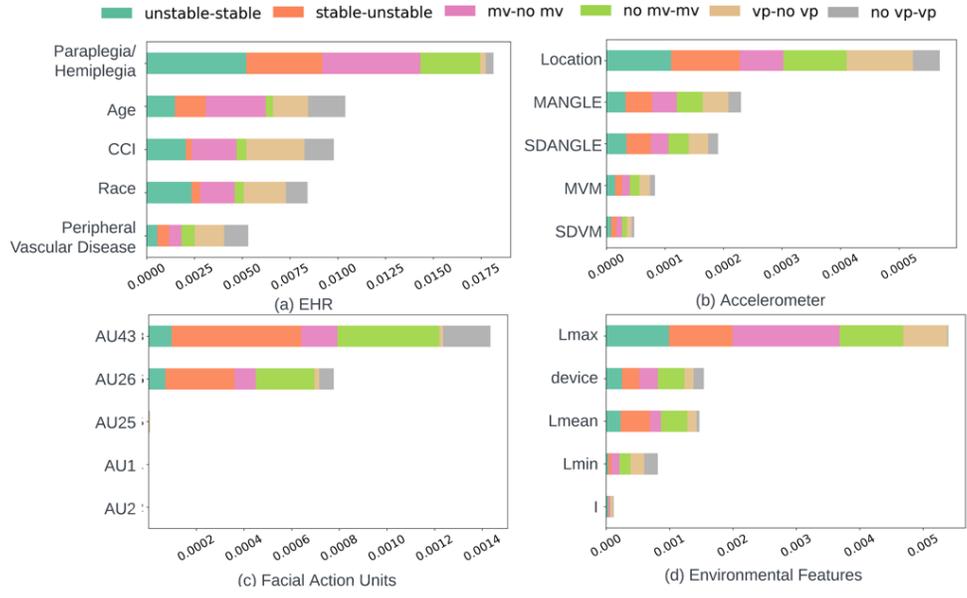

A

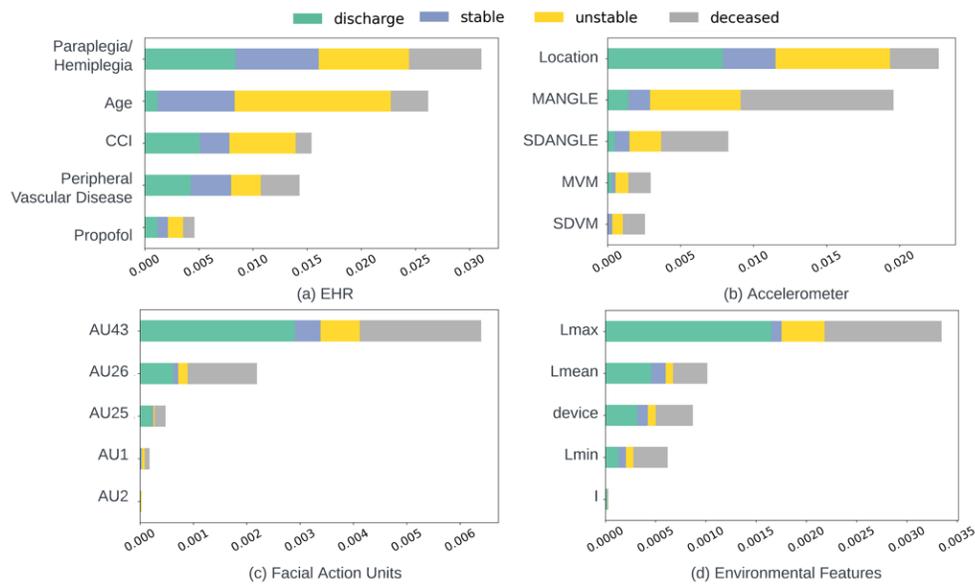

B

Fig. 2. The bar plots A and B show the top five important features across $Y_{trans}$ and $Y_{status}$ for different modalities in the experiment of all the modalities being involved. Each color denotes a different outcome, reflecting the average absolute importance of each feature. A (a) and B (a) show the five most important features in the EHR data. Paraplegia/Hemiplegia, age, and CCI were the strongest predictors in most of the tasks. A (b) and B (b) highlight the five most important features among all the accelerometer features in the two main tasks. The location of the sensor and the mean of the angle between x and vector magnitude acted as the most important feature across all the sub-tasks. A (c) and B (c) showcased the feature importance of face data. AU43 (eyes close) and AU26 (Jaw drop) dominated the prediction. (d) Both A and B illustrate the top five features in environmental features, where the sound pressure level features contributed the most compared with light intensity.

Abbreviations: CCI: Charlson Comorbidity Index

## IV. Discussion

The objective of this study was to build a multimodal dataset and a deep learning model to predict acuity status, transitions in acuity and the need for life-sustaining therapies. Four data modalities, namely EHR, accelerometry, facial AU and environmental data were used for this task. A novel multimodal feature fusion strategy was developed to fuse the four data modalities. A masking scheme was developed to generalize the model to scenarios where one or more data modalities might be missing.

The experiments revealed that EHR, facial AUs, and accelerometer data were more predictive of the transition outcomes in this study. The integrated gradient analysis further highlighted the feature importance of these three modalities. Notably, the top five important features from the face AUs and EHR data in both transitions and status prediction aligned with the findings of previously existing literature in this domain [3, 7]. For EHR data, as seen in both Fig. 2 A and B, patients' demographics, such as age and race, contributed highly to the transition prediction. Race could relate to the patient's socioeconomic status which further underlay insurance status, access to healthcare, or community health metrics. A cohort study has emphasized the impact of race as a potential proxy for readmissions [26]. Future research should consider incorporating such demographic variables to disentangle the effects of race from these confounding factors, thereby improving the specificity of predictive models and ensuring AI fairness in healthcare analytics.

Paraplegia/Hemiplegia is a common comorbidity that represents severe neurological deficits resulting from spinal cord injuries or cerebrovascular accidents. Similarly, PVD (Peripheral Vascular Disease) reflects systemic atherosclerotic disease affecting peripheral arteries. These complications necessitate close monitoring and aggressive treatment strategies, thereby increasing the patient's acuity. In sub-plot (b) of both Fig. 2 A and B, the top-ranked feature of accelerometer data is the mean of the angle between the x-axis and vector magnitude, followed by the standard deviation of vector angle. These two features could reflect patient mobility by showing the orientation of the movement. Notably, the orientation information was more predictive than the magnitude of the movement, as it directly correlates with the directional characteristics of movement, enabling detailed understanding of posture changes and movement patterns. For facial AUs, AU43 (eyes close) and AU26 (Jaw drop) were highly ranked in both prediction tasks. AU43 is used in the Prkachin Solomon Pain Index (PSPI) to assess pain levels [20]. EHR features, accelerometer features, and face AUs played critical roles in predicting both transitions and status prediction, as shown in Table V and Table IV. However, the environmental data, which comprised light intensity and sound pressure level, did not significantly improve the model's prediction performance. This could result from the weak correlation between environmental features and acuity. Environmental factors have been proven to contribute to the onset of delirium [27]. However, for acuity, the existing literature has only suggested that sleep disruption caused by environmental factors may contribute to an increase in patient acuity [28]. There was no documented evidence directly linking the environment to changes in acuity.

Although our multimodal approach has shown promising performance, this work has some limitations. Our study was limited to 310 patients, all recruited at one center, which restricts the generalizability of the models. Furthermore, all four data modalities were not available for every patient. Specifically, during a patient's stay in the ICU, EHR data was continuously recorded until transfer or discharge. In contrast, data from other modalities were only collected for seven days or until the patient was transferred or discharged from the ICU.

Consequently, for patients with extended ICU stays, the proportion of multimodal data is significantly reduced, impacting the overall integration and effectiveness of the multimodal approach. We used statistical features for face AUs, accelerometers, and environmental data. Due to the relatively large time window of four hours employed in our work, statistical features tended to average out significant information. This averaging effect can lead to a loss of important dynamic patterns that might be critical for accurate predictions and assessments. Additionally, the temporal effects caused by creating time windows have been proven to affect the model performance [29]. Employing large time windows may obscure essential temporal dynamics, potentially diminishing the model's predictive accuracy.

We are committed to patient data security and privacy, following all regulations and privacy rules. The raw data of different modalities was stored and organized on a dedicated server isolated from the internet, accessible only through the University of Florida health VPN. During data processing, all the raw data was de-identified to ensure the researchers had no access to any identifying patient information.

In future work, we will focus on refining the multimodal model architecture to directly process raw data inputs instead of relying on statistical features. This approach will be tested on a larger dataset and incorporate additional modalities, such as depth images widely used in patients' mobility assessments [30], to further enhance the model's predictive capabilities. Additionally, a real-time multimodal model for ICU outcomes prediction will be developed and employed to liberate manpower further and benefit the decision-making process in clinical settings.

## V. Conclusion

In this study, we introduced MANGO, the first multimodal attention-based deep learning model designed to predict transitions in patients' acuity status and the need for life-sustaining therapies and to classify patients' acuity status with four different modalities in the ICU. By leveraging the proposed multimodal feature fusion method and masked multi-head self-attention mechanism, we addressed the limitations inherent in single-modality data approaches and leveraged the complement information from multiple data sources. Four modalities were used in this study: EHR, accelerometry data, changes in patients' faces obtained in the form of facial AUs extracted from video data, and environmental data, which included light and noise levels in the ICU. Our novel multimodal feature fusion strategy demonstrated the ability to maintain robust performance despite missing modalities. In the future, we will enlarge our cohort and introduce more complex modalities, such as depth images and other sensor data sources, with more advanced multimodal training methods.

## IV. Acknowledgement

A.B, P.R., and T.O.B. were supported by NIH/NINDS R01 NS120924, NIH/NIBIB R01 EB029699. P.R. was also supported by NSF CAREER 1750192.

# References

[1] R. Neal *et al.*, "Is increased time to diagnosis and treatment in symptomatic cancer associated with poorer outcomes? Systematic review," *British journal of cancer,* vol. 112, no. 1, pp. S92-S107, 2015.

[2] S. Nerella *et al.*, "Transformers and large language models in healthcare: A review," *Artificial Intelligence in Medicine,* p. 102900, 2024.

[3] M. Contreras *et al.*, "APRICOT-Mamba: Acuity Prediction in Intensive Care Unit (ICU): Development and Validation of a Stability, Transitions, and Life-Sustaining Therapies Prediction Model," 2023.

[4] B. Shickel, T. J. Loftus, L. Adhikari, T. Ozrazgat-Baslanti, A. Bihorac, and P. Rashidi, "DeepSOFA: a continuous acuity score for critically ill patients using clinically interpretable deep learning," *Scientific reports,* vol. 9, no. 1, p. 1879, 2019.

[5] T. Wenham and A. Pittard, "Intensive care unit environment," *Continuing Education in Anaesthesia, Critical Care & Pain,* vol. 9, no. 6, pp. 178-183, 2009.

[6] J. Sena *et al.*, "Wearable sensors in patient acuity assessment in critical care," *Frontiers in Neurology,* vol. 15, p. 1386728, 2024.

[7] S. Nerella *et al.*, "Detecting Visual Cues in the Intensive Care Unit and Association with Patient Clinical Status," *arXiv preprint arXiv:2311.00565,* 2023.

[8] T. Shaik, X. Tao, L. Li, H. Xie, and J. D. Velásquez, "A survey of multimodal information fusion for smart healthcare: Mapping the journey from data to wisdom," *Information Fusion,* p. 102040, 2023.

[9] Q. Cai, H. Wang, Z. Li, and X. Liu, "A survey on multimodal data-driven smart healthcare systems: approaches and applications," *IEEE Access,* vol. 7, pp. 133583-133599, 2019.

[10] Y. Ma *et al.*, "Global Contrastive Training for Multimodal Electronic Health Records with Language Supervision," *arXiv preprint arXiv:2404.06723,* 2024.

[11] X. Li, J. Gu, Z. Wang, Y. Yuan, B. Du, and F. He, "XAI for In-hospital Mortality Prediction via Multimodal ICU Data," *arXiv preprint arXiv:2312.17624,* 2023.

[12] P. Ekman and W. V. Friesen, "Facial action coding system," *Environmental Psychology & Nonverbal Behavior,* 1978.

[13] Y. Ren *et al.*, "Computable Phenotypes of Patient Acuity in the Intensive Care Unit," *arXiv preprint arXiv:2005.05163,* 2020.

[14] M. Contreras *et al.*, "APRICOT-Mamba: Acuity Prediction in Intensive Care Unit (ICU): Development and Validation of a Stability, Transitions, and Life-Sustaining Therapies Prediction Model," *arXiv preprint arXiv:2311.02026,* 2023.

[15] M. Kheirkhahan *et al.*, "A smartwatch-based framework for real-time and online assessment and mobility monitoring," *Journal of biomedical informatics,* vol. 89, pp. 29-40, 2019.

[16] Z. Liu *et al.*, "Swin transformer: Hierarchical vision transformer using shifted windows," in *Proceedings of the IEEE/CVF International Conference on Computer Vision*, 2021, pp. 10012-10022.

[17] X. Zhang et al., "A high-resolution spontaneous 3d dynamic facial expression database," in *2013 10th IEEE international conference and workshops on automatic face and gesture recognition (FG)*, 2013: IEEE, pp. 1-6.

[18] S. M. Mavadati, M. H. Mahoor, K. Bartlett, P. Trinh, and J. F. Cohn, "Disfa: A spontaneous facial action intensity database," *IEEE Transactions on Affective Computing,* vol. 4, no. 2, pp. 151-160, 2013.

[19] P. Lucey, J. F. Cohn, K. M. Prkachin, P. E. Solomon, and I. Matthews, "Painful data: The UNBC-McMaster shoulder pain expression archive database," in *2011 IEEE International Conference on Automatic Face & Gesture Recognition (FG)*, 2011: IEEE, pp. 57-64.

[20] S. Nerella, J. Cupka, M. Ruppert, P. Tighe, A. Bihorac, and P. Rashidi, "Pain action unit detection in critically ill patients," in *2021 IEEE 45th Annual Computers, Software, and Applications Conference (COMPSAC)*, 2021: IEEE, pp. 645-651.

[21] J. Redmon, S. Divvala, R. Girshick, and A. Farhadi, "You only look once: Unified, real-time object detection," in *Proceedings of the IEEE conference on computer vision and pattern recognition*, 2016, pp. 779-788.

[22] S. Bandyopadhyay et al., "Predicting risk of delirium from ambient noise and light information in the ICU," *arXiv preprint arXiv:2303.06253,* 2023.

[23] O. M. Buxton and E. Marcelli, "Short and long sleep are positively associated with obesity, diabetes, hypertension, and cardiovascular disease among adults in the United States," *Social science & medicine,* vol. 71, no. 5, pp. 1027-1036, 2010.

[24] A. Vaswani et al., "Attention is all you need," *Advances in neural information processing systems,* vol. 30, 2017.

[25] M. Sundararajan, A. Taly, and Q. Yan, "Axiomatic attribution for deep networks," in *International conference on machine learning*, 2017: PMLR, pp. 3319-3328.

[26] J. Meddings et al., "The impact of disability and social determinants of health on condition-specific readmissions beyond Medicare risk adjustments: a cohort study," *Journal of general internal medicine,* vol. 32, pp. 71-80, 2017.

[27] T. D. Girard, P. P. Pandharipande, and E. W. Ely, "Delirium in the intensive care unit," *Critical care,* vol. 12, pp. 1-9, 2008.

[28] M. A. Pisani, R. S. Friese, B. K. Gehlbach, R. J. Schwab, G. L. Weinhouse, and S. F. Jones, "Sleep in the intensive care unit," *American journal of respiratory and critical care medicine,* vol. 191, no. 7, pp. 731-738, 2015.

[29] W. Liu, Z. He, and X. Huang, "Time Matters: Examine Temporal Effects on Biomedical Language Models," *arXiv preprint arXiv:2407.17638,* 2024.

[30] M. D. Hashem, A. M. Parker, and D. M. Needham, "Early mobilization and rehabilitation of patients who are critically ill," *Chest,* vol. 150, no. 3, pp. 722-731, 2016.